\documentclass[iop]{emulateapj}
\usepackage{natbib,graphicx,amsmath,xspace}

\begin{document}

\title{Calculating Time Lags From Unevenly-Sampled Light Curves}

\author{A. Zoghbi$^{1,2}$, C. Reynolds$^{1,2}$, E. M. Cackett$^3$}
\affil{$^1$Department of Astronomy, University of Maryland, College Park, MD 20742-2421, USA}
\affil{$^2$Joint Space-Science Institute (JSI), College Park, MD 20742-2421, USA}
\affil{$^3$Department of Physics \& Astronomy, Wayne State University, 666 W. Hancock St, Detroit, MI 48201}
\email{azoghbi@astro.umd.edu}

\begin{abstract}
Timing techniques offer powerful tools to study dynamical astrophysical phenomena. In the X-ray band, they offer the potential of probing accretion physics down to the event horizon. Recent work has used frequency and energy-dependent time lags as a tool for studying relativistic reverberation around the black holes in several Seyfert galaxies. This was achieved thanks to the evenly-sampled light curves obtained using XMM-Newton. Continuous-sampled data is however not always available and standard Fourier techniques are not applicable. Here, building on the work of \cite{2010MNRAS.403..196M}, we discuss and use a maximum likelihood method to obtain frequency-dependent lags that takes into account light curve gaps. Instead of calculating the lag directly, the method estimates the most likely lag values at a particular frequency given two observed light curves. We use Monte Carlo simulations to assess the method's applicability, and use it to obtain lag-energy spectra from \emph{Suzaku} data for two objects, NGC 4151 and MCG-5-23-16, that had previously shown signatures of iron K reverberation. The lags obtained are consistent with those calculated using standard methods using XMM-Newton data.
\end{abstract}

\section{Introduction}
Variability is ubiquitous in astrophysical phenomena. The observed fluxes are seen to vary on different time-scales for the different classes of objects. From milliseconds in neutron stars, stellar mass black holes and Gamma Ray bursts, to progressively increasing time-scales for the different phases of planetary, stellar and galaxy evolution. Variability has been invaluable in understanding the dynamics of many systems that is otherwise observationally inaccessible.

Although a lot of the methodology relies on time-domain analysis, frequency-domain techniques remain the standard tool in characterizing time-scale dependency and (quasi-)periodicity when good data coverage is available. Stellar mass black holes and neutron star LMXBs in particular show a rich phenomenology in the frequency domain, that is very tied to state transitions  that have distinct spectroscopic signatures \citep{2000ARA&A..38..717V,2006ARA&A..44...49R}.

In addition to estimating the power spectral density (PSD) in the frequency domain for a single light curve, other measures exist when multiple, simultaneous (e.g. at different bands or energies) light curves exist. The cross spectrum gives a measure of the combined variability power in two light curves, the coherence measures the fraction of one light curve that can be predicted from the other \citep{1997ApJ...474L..43V}, and the phase lag gives the relative delay (in units of radians, between $-\pi$ and $\pi$) between the two light curves as a function of frequency \citep{1989Natur.342..773M}. The phase lag is converted into a time lag (in units of seconds) by dividing the phase lag by the angular frequency of the measurement.

In AGN X-ray studies that motivated this work, frequency-domain techniques have been used extensively to characterize the broadband variability \citep{1995MNRAS.273..923P,2002MNRAS.332..231U,2003MNRAS.345.1271V,2004MNRAS.348..783M}. Periodicity is generally not seen \cite[see][ for the exception]{2008Natur.455..369G}, and the power spectra are characterized with a power-law of index $\sim2$ at high frequencies that breaks or bends to $\sim1$ at a characteristic  frequency that appears to scale with mass \citep{2006Natur.444..730M}. The standard tools in this case is the Fast Fourier  Transform (FFT), which, for a continuously-sampled time series, gives a set of complex numbers at specific frequencies. The periodogram, which estimates the power spectrum, is the squared amplitude of these complex transforms. 

Inter-band time delays in the standard X-ray bandpass ($0.3-20$ keV) in AGN have just started to be explored in detail. Low frequency hard lags have been seen in early XMM-Newton observations \citep{2001ApJ...554L.133P,2007MNRAS.382..985M,2008MNRAS.388..211A}, where hard bands lag softer bands and the lag magnitude depends on the separation of energies, similar to that in stellar mass black holes \citep{1999ApJ...510..874N,2001MNRAS.327..799K}. More recently, high frequency lags have also been seen \citep[e.g.][]{2009Natur.459..540F,2010MNRAS.401.2419Z,2013MNRAS.431.2441D,2013ApJ...764L...9C}. In this case, the lag can be soft or hard depending on the selected energies, but it is distinguished from the low frequency lags by its energy dependence. The shape of the lag-energy spectra, which gives a measure of the inter-band delays as a function of energy at a particular Fourier frequency, appears to be closely related to the spectroscopic components in a standard spectrum \citep{2011MNRAS.412...59Z,2013MNRAS.430.1408K,2013ApJ...767..121Z}. The fact that the $1-3$ keV band leads both the $<1$ keV and $>3$ keV at high frequencies points to a reflection origin, with relativistic reflection being the most plausible explanation. In this case, the reflection spectrum, matched by the lag-energy spectrum, is produced within $\sim 10$ gravitational radii from the event horizon of the black hole, and reverberation is produced when the reflecting medium responds to the fast variations of the illuminating source, providing a powerful tool to probe these environments \citep{1999ApJ...514..164R}.

Similar to power spectra, time lags in these cases are calculated from the FFT (see sec. \ref{sec:std_fft}) for continuously-sampled light curves. The statistical properties of lag measurements in this case are discussed in \cite{1999ApJ...510..874N}. Extending reverberation studies beyond XMM-Newton data is not possible using the standard Fourier techniques because of the inherent non-continuous sampling forced by the low earth orbits of other observatories like \emph{Suzaku}, NuSTAR and AstroH. The lowest frequency probed with the standard Fourier techniques are those associated with the orbital period of the satellite, which for the case of low-earth orbits is higher than the frequency of the interesting reverberation in AGN. \cite{2010MNRAS.403..196M} introduced a method based on likelihood maximization that directly fits for frequency-dependent time lag \citep[along with the power spectrum, based on the work of][]{1998PhRvD..57.2117B}. In this work, we explicitly discuss it in detail, assessing its applicability using Monte Carlo simulations. Then, we apply it to \emph{Suzaku} observations of two objects, NGC 4151 and MCG-5-23-16, that had previously shown relativistic reverberation delays in the iron K band. We start section \ref{sec:std_fft} of this article by reviewing the standard Fourier techniques for both power spectra and time lags. In section \ref{sec:likelihood}, we describe the formalism of the likelihood method. Section \ref{sec:sims} discusses the detailed Monte Carlo simulations of the applicability of the method. The application of the method to \emph{Suzaku} archival observations of NGC 4151 and MCG-5-23-16 is presented in section \ref{sec:applications}.

\section{Standard Fourier Techniques}\label{sec:std_fft}
In this section, we briefly review the standard techniques based on the Fourier transform, which are commonly used with evenly-sampled light curves. The power spectral density $\mathcal{P}(f)$ is a property of the stochastic process producing the variability, and it gives a measure of the variability power as a function of temporal frequency $f$. It is estimated by calculating the periodogram $I$. If the observed data is in the form of a vector $\boldsymbol{x}$ of length $N$ that gives the count rates at times $t_i=i\Delta t$, where $i$ takes the integral values $1,2,..., N$ and $\Delta t$ is the time bin size, the periodogram $I(f)$ is given by the squared amplitude of Discrete Fourier Transform of $\boldsymbol{x}$:
\begin{equation}
I(f_j) = A\left|\sum\limits_{i=1}^{N}x_i e^{i2\pi f_jt_i}\right|^2
\end{equation}
where $f_j = j/N\Delta t$ with $j=1,2,...,N/2$. A is a normalization factor, which we take in this work to be $A=2\Delta t/N$ \citep{2003MNRAS.345.1271V}. The periodogram $I$ itself is an {\it inconsistent} estimator of $\mathcal{P}$, where its standard deviation at frequency $f$ is equal to its value \citep{nla.cat-vn2888327}. The variance is reduced significantly if several frequencies are grouped together \citep[e.g.][]{1993MNRAS.261..612P}.

Let us consider a second light curve $\boldsymbol{y}$ which gives the count rate at the same time intervals $t_i = i\Delta t$ but in another energy band. The cross spectrum can be estimated as $C(f) = X^{\ast}(f)Y(f)$, where $X$ and $Y$ are the Fourier transform of $\boldsymbol{x}$ and $\boldsymbol{y}$ respectively, and the $X^{\ast}$ is the complex conjugate of $X$. The cross spectrum is a complex number. Its amplitude is usually expressed in the form of the coherence function $\gamma^2(f)=|\langle C\rangle|^2/(\langle|X|^2\rangle\langle|Y|^2\rangle)$ \citep{1997ApJ...474L..43V}, where the angle brackets denote averaging. The phase of the complex cross spectrum gives the phase lag between the two light curves \citep{1989Natur.342..773M,1999ApJ...510..874N}:

\begin{alignat}{1}
\phi(f) = {\rm arg}[C(f)]&\hspace{0.5cm}
\end{alignat}
The time lag $\tau(f)$ is then obtained by dividing by $2\pi f$, so that $\tau(f) = \phi(f)/2\pi f$. $\tau(f)$ gives measure of time delay between $\boldsymbol{x}$ and $\boldsymbol{y}$ as a function of frequency (or variability time-scale).

The above calculations require the light curve to be evenly-sampled so the Fourier transform can be utilized. If this is not the case, other techniques are needed. The following section discusses the method of using the likelihood function to directly fit for the best estimates for the power and cross spectra as well as the phase/time lags directly.

\section{Likelihood Analysis}\label{sec:likelihood}
The principle idea behind the method was first presented in the context of X-ray light curves by \cite{2010MNRAS.403..196M}. Here, we expand it and show explicitly how the method works and perform Monte Carlo simulation to assess its applicability.
The method fits for the most likely variability powers and time lags given the observed data. Starting with a model for the power and time lags (which can be of a functional form such as a power-law, or parameterized with the values of power and time lags in pre-defined frequency bins), a likelihood function that compares the model to the data is constructed (by comparing the auto- and cross-correlations of the data with those expected from the model), and the best estimates are obtained by maximizing this likelihood function.
We start in section \ref{likelihood_psd} by applying it to estimating the power spectrum, then in section \ref{likelihood_lag}, with a simple extension, we use the method to estimate the cross power spectrum and the time lag as a function of Fourier frequency.

\subsection{Power Spectrum Estimate}\label{likelihood_psd}
As before, the light curve is taken to be $\boldsymbol{x}$ with values $x_i$ for $i=1,...,N$, but now $t_i \not= i\Delta t$. Following \cite{1998PhRvD..57.2117B} (although that work is for 2d CMB data), each $x_i$ is the sum of the contribution from the signal $s_i$ and noise $n_i$. The noise is assumed Gaussian, which is almost always the case given that each $x_i$ results from binning measurements obtained at a sampling smaller than $\Delta t$. So the observed light curve is:

\begin{equation}\label{eqn:x_s_n}
\boldsymbol{x} = \boldsymbol{s} + \boldsymbol{n}
\end{equation}
with a correlation matrix given by:
\begin{equation}\label{eqn:def_Cx}
\langle x_i x_j \rangle \equiv \mathsf{C}_x = \mathsf{C}_s + \mathsf{C}_n
\end{equation}
where $\mathsf{C}_s = \langle s_i s_j\rangle$ is the source signal correlation matrix and $\mathsf{C}_n = \langle n_i n_j\rangle$ is the noise correlation matrix. The angle brackets indicate ensemble average and we have assumed that the source noise components are independent. Because the measurement errors in light curves are independent, $\mathsf{C}_n$ is diagonal with entries $n_i n_i$, $i=1,...,N$. In general, if the observations have correlated noise, they can be easily incorporated here by adding non-diagonal elements to $\mathsf{C}_n$.

$\mathsf{C}_s$ is unknown, and its components $c_{ij}$ defined by $\tau=t_j-t_i$ are related to the underlying power spectrum through the autocorrelation function $\mathcal{A}(\tau)$.

\begin{equation}\label{eqn:po_spec}
\langle s(t)s(t+\tau)\rangle = \mathcal{A}(\tau) = \int_{-\infty}^{+\infty} \mathcal{P}(f){\rm cos}(2\pi f \tau) df
\end{equation}
using the relation that the autocorrelation is the Fourier transform of the power spectrum, and for real functions only the cosine term is included. Now starting from $\mathcal{P}(f)$ that depends on a number of parameters $\boldsymbol{a_p}$ (of length $n_p$ say) to be found, we construct $\mathsf{C}_s$ and calculate the likelihood functions for those parameter:
\begin{equation}\label{likelihood_eqn}
\mathcal{L}(\boldsymbol{a_p}) = (2\pi)^{-N/2} |\mathsf{C}_x|^{-1/2} {\rm exp} \left[ -\frac{1}{2} \boldsymbol{x}^T \mathsf{C}_{x}^{-1} \boldsymbol{x} \right]
\end{equation}
where the dependence on $\boldsymbol{a_p}$ is in $\mathsf{C}_x$ through its dependence on $\mathcal{P}$, and $\boldsymbol{x}^T$ is the transpose of $\boldsymbol{x}$. Thus, $\mathsf{C}_x$ is calculated from the model and $\boldsymbol{x}$ is the data vector. The procedure now is to select a model $\mathcal{P}(f)$ and fit for the parameters $\boldsymbol{a_p}$ that maximize the likelihood function in equation \ref{likelihood_eqn}. $\mathcal{P}$ can be taken to be a power-law or a broken power-law function of $f$. Alternatively, we can fit the band powers directly, taking the powers in pre-defined frequency bands as the parameters $\boldsymbol{a_p}$. This is the best option when the intrinsic shape is unknown, which might not be the case for $\mathcal{P}$, but is certainly the case for frequency-dependent lags $\tau(f)$.

The standard is to maximize $log(\mathcal{L})$ instead of $\mathcal{L}$, and because of the functional form of the likelihood, the gradient and the second derivatives of the likelihood can be calculated \citep{1998PhRvD..57.2117B}. An iterative quadratic approximation is then used to find the maximum likelihood. The structure of the log-likelihood function is relatively smooth to converge within a few iterations.

\subsection{Time Lag Estimate}\label{likelihood_lag}
Extending the previous formalism to include time lags is straight forward. Now we have another light curve $\boldsymbol{y} = \boldsymbol{r}+\boldsymbol{n}_y$, that represents the count rates in a different band for example. $\boldsymbol{y}$ can be appended to the vector $\boldsymbol{x}$ to give an {\it augmented} data vector $\boldsymbol{\tilde{x}}=\left(\begin{array}{l} \boldsymbol{x} \\ \boldsymbol{y} \end{array}\right)$ \citep{1992ApJ...398..169R}. The covariance matrix $\mathsf{\tilde{C}}$ of the new data vector is:

\begin{equation}
\mathsf{\tilde{C}} = \left( \begin{array}{cc}
\mathsf{C}_x & \mathsf{C}_{xy} \\
\mathsf{C}_{xy}^T & \mathsf{C}_{y} \\
\end{array} \right)
\end{equation}
where $\mathsf{C}_y$ is the covariance matrix of the second light curve $\boldsymbol{y}$ defined in a similar way to equation \ref{eqn:def_Cx}, with $\mathsf{C}_r$ being in general different from $\mathsf{C}_s$. The matrix $\mathsf{C}_{xy}$ is the cross-covariance matrix of $\langle x_i y_j \rangle$. The noise components of $\boldsymbol{x}$ and $\boldsymbol{y}$ are assumed independent because the two light curves are produced by independent events in the two bands. The noise components are also independent of the two light curves so that $\langle n_i x_j\rangle = 0$, $\langle n_i y_j\rangle = 0$ etc. Therefore $\mathsf{C}_{xy} = \langle s_i r_j\rangle$.

The cross-covariance is related to the cross-power spectrum and the phase lag through the cross-correlation function $\mathcal{X}(\tau)$:

\begin{equation}\label{eqn:cx_spec}
\langle s(t)r(t+\tau)\rangle = \mathcal{X}(\tau) = \int_{-\infty}^{+\infty} \mathcal{C}(f){\rm cos}(2\pi f \tau + \phi(\tau) ) df
\end{equation}

The parameters we are interested in are now in the matrix $\mathsf{\tilde{C}}$. The likelihood equation is similar to equation \ref{likelihood_eqn}, replacing $\boldsymbol{x}$ and $\mathsf{C}$ with $\boldsymbol{\tilde{x}}$ and $\mathsf{\tilde{C}}$ respectively. Again, we use a pre-selected model (e.g. power-law), or choose the powers and lags in pre-defined frequency bins as the parameters of interest. In this work, we choose the latter parameterization. If $n_{B}$ is the number of frequency bins, then we have $n_p=4n_{B}$ parameters: the powers for each light curve, the cross powers and the phase lags. For this case, the maximization procedure starts with obtaining the PSD values for individual light curves first, then for the cross power and phase lags.

In practice, there are also other effects that need to be considered. Aliasing is a consequence of the fact that power cannot be calculated beyond the Nyquist frequency $f_N=1/2\Delta t$. The result is that the measured powers at a frequency $f$ contain also contribution from its aliases above $f_N$. Fortunately however, X-ray light curves generally have power-law PSDs, so the power above $f_N$ is small, and also the fact $\Delta t$ is a width of a bin not the actual sampling time. The binning process is equivalent to convolving the light curve with a binning window $b(t)=1/\Delta t$ for $-\Delta t/2<t<\Delta t/2$ and $0$ otherwise. The result is that the $\mathcal{P}$ is multiplied by the Fourier transform of $b(t)$, which is: $sinc^2(\pi f \Delta t)$ \citep[e.g.][]{1989tns..conf...27V}.
Red noise leak is another problem and it is the result of the finite length of the observation. $\mathcal{P}$ in this case is convolved with the Fourier transform of the window function, and mainly causes power below the lowest measured frequency ($f_{min}=1/T$, where $T$ is the length of the observation) to leak into frequencies above $f_{min}$. One can explicitly include the convolution of the window in equations \ref{eqn:po_spec} and \ref{eqn:cx_spec}. However, we found that it is computationally easier to include the additional power below $f_{min}$ in the fit by extending the lowest boundary of the lowest frequency bin to values smaller than $f_{min}$, this was found to correct for the power biases (see sections \ref{sec:psd_sim}). Extending the first bin to frequencies less than $f_{min}$ assumes that the power below $f_{min}$ does not change significantly, which is a reasonable assumption given that the psd in almost all cases is a smooth power-law.

\subsection{Estimating Uncertainties}\label{sec:uncertainties}
As discussed in \cite{2010MNRAS.403..196M}, the uncertainties can be estimated, as part of the fitting procedure, by calculating the Fisher matrix, which is related to the second derivative of the log-likelihood \citep[see detailed related discussion in][]{1997ApJ...480...22T}. The Fisher matrix basically measures how fast on average does the likelihood function fall off around the its maximum. When the best fit is found, an estimate of the covariance matrix of the parameters is given by the inverse of the Fisher matrix. The variance of the estimates parameters are the diagonal elements of this covariance matrix. These estimates are however only a lower limit on the uncertainties when the off-diagonal values are not small (i.e. parameters are correlated).

The alternative is to step through the parameters, taking the $68\%$ uncertainty as the value that changes $-2log(\mathcal{L}/\mathcal{L}_{max})$ by 1 \citep{2010MNRAS.403..196M}. Another approach involves using Monte Carlo Markov Chain (MCMC) to map the full probability space, obtaining probability distributions for the parameters directly. The uncertainties quoted in this work, unless stated otherwise, are the result of stepping through each parameter individually, allowing the rest to change, and taking the error as the value that changes the value of $-2log(\mathcal{L}/\mathcal{L}_{max})$ by 1. This choice works when the number of parameters to be fitted is small ( $n_p<\sim20$, so stepping through parameters is computationally feasible relatively quickly). If the number is large, the best option is to use MCMC to obtain uncertainties.

\section{Simulations}\label{sec:sims}
In order to test the above method, we simulate light curves with known underlying power-spectra and time delays, introduce gaps, and explore how well they can be recovered. Starting with a functional form for the $\mathcal{P}$, we randomize the amplitude and the phase then inverse Fourier transform to obtain one light curve realization \citep{1995A&A...300..707T}. When a second light curve is needed, we shift the phase by the desired amount before performing the inverse Fourier transform. This assumes unity coherence. When fitting real data, the coherence can be estimated from the cross spectrum and the individual power spectra. Poisson noise is added to all light curves.
\begin{figure}
\centering
 \includegraphics[width=210pt,clip ]{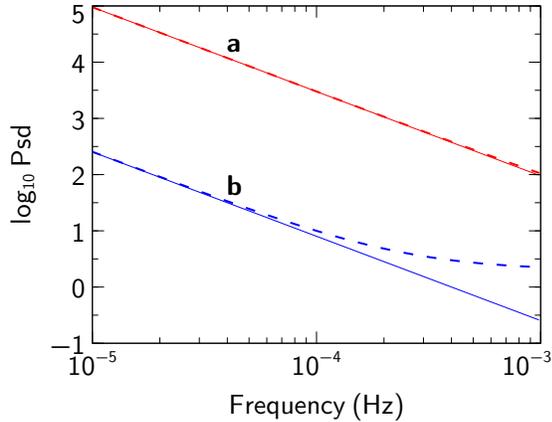}
\caption{The two cases of base power-spectra used in the simulations. Solid lines represent the underlying generating spectrum. Dashed lines include the effect of Poisson noise. Case 1 corresponds to a bright variable ($35\%$ rms) source and case 2 corresponds to a relatively fainter and less variable source ($10\%$ rms).}
\label{fig:psds}
\end{figure}

\begin{figure}
\centering
 \includegraphics[width=230pt,clip ]{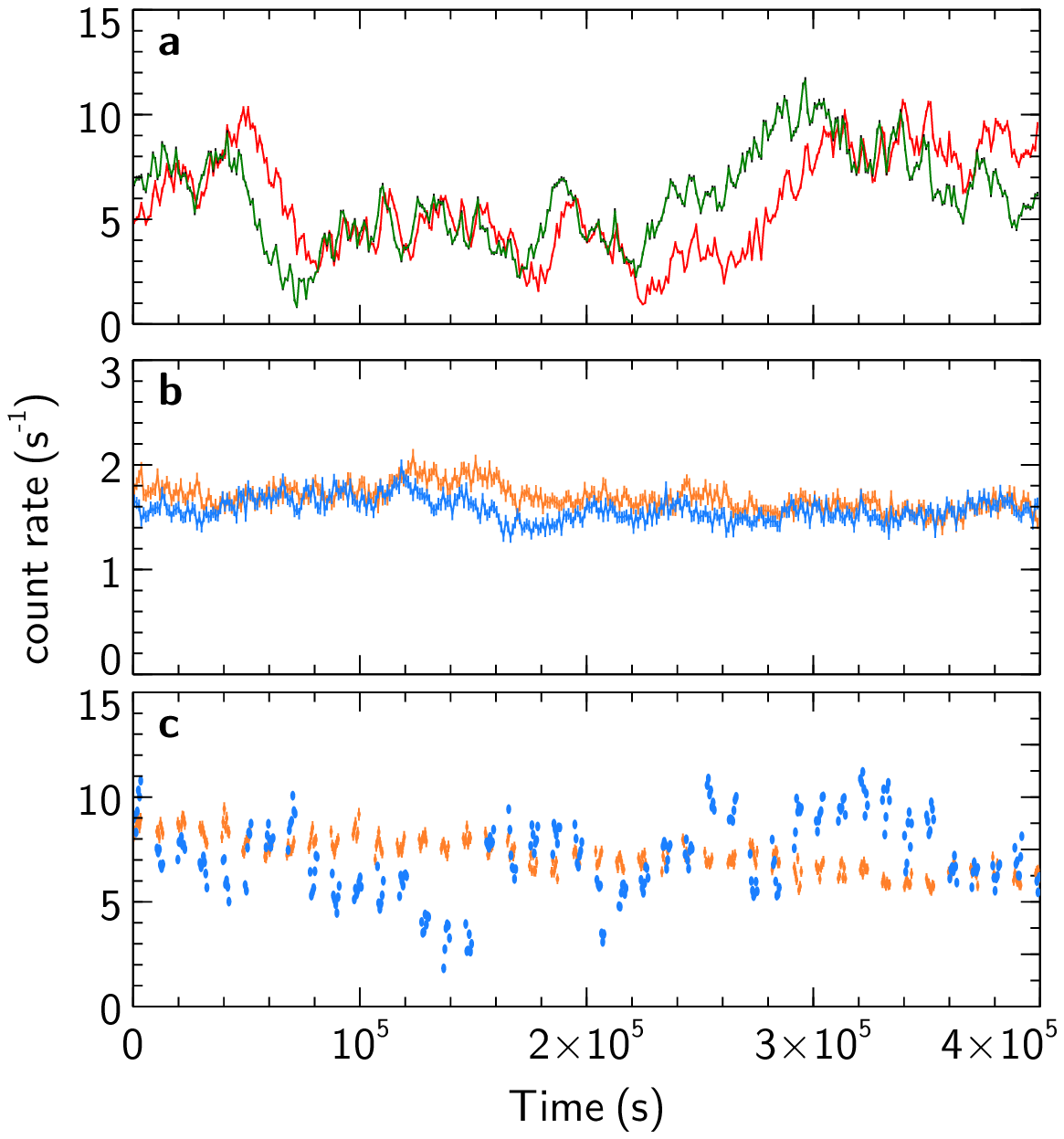}
\caption{Typical light curve realizations from cases 1 (panel a) and 2 (panel b) in Fig.\ref{fig:psds}. In each case, the second light curve is lagged by 1 radians with respect to first. Panel c shows typical light curves with gaps for the two cases. The y-axis is similar to panels a and b.}
\label{fig:lc}
\end{figure}

\begin{figure}
\centering
 \includegraphics[width=250pt,clip ]{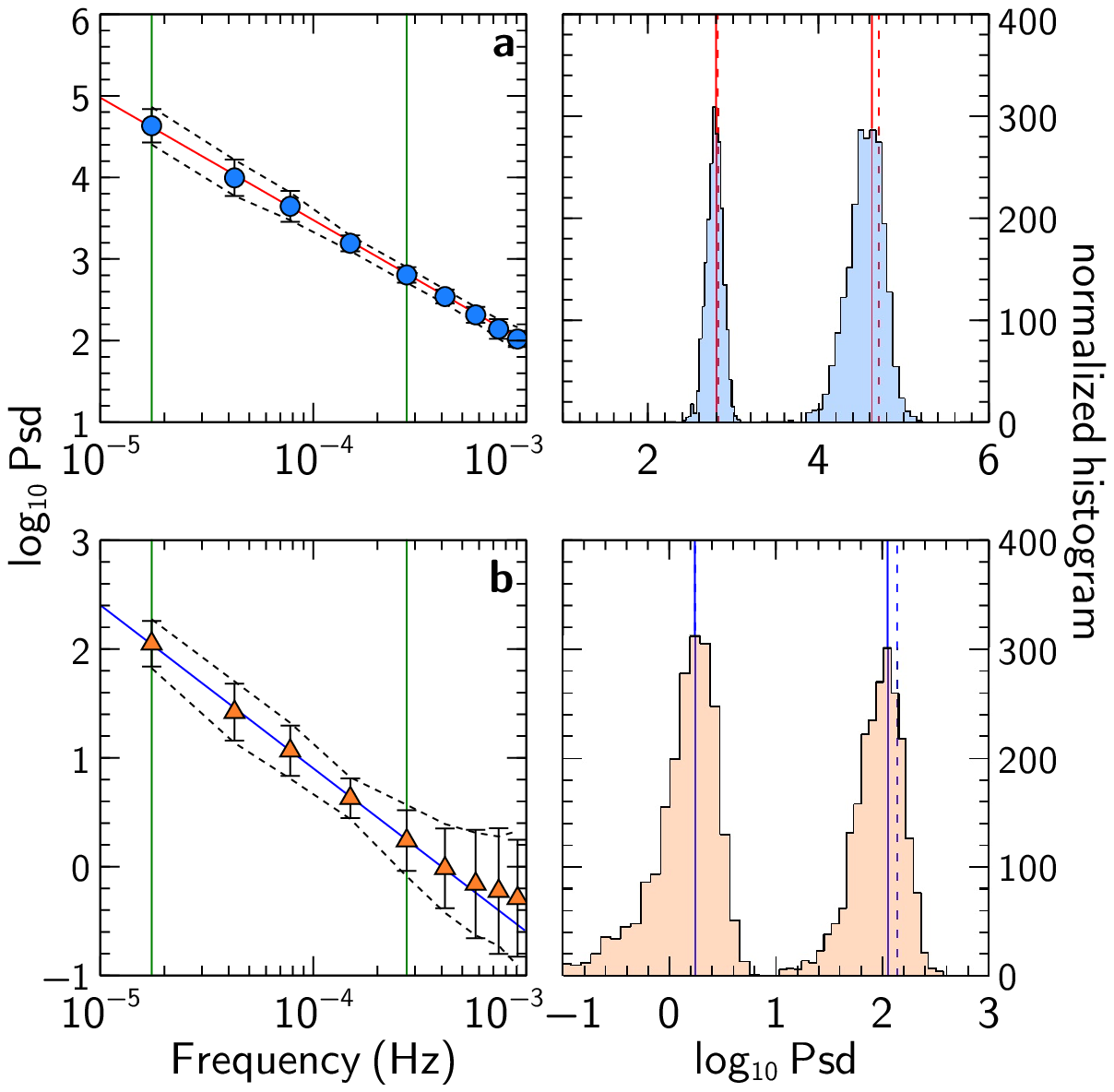}
\caption{{\it Left:} The average estimated PSD for the high (a) and low-rms (b) cases without including gaps. {\it Right: } The histogram of the values for two selected frequency bands (marked with vertical lines in the left panel ). For each of the high and low-rms cases, more than 2000 separate light curves are simulated. For each one, the power is estimated at nine frequency bins. The means of the resulting distributions are plotted in the left panel. Their errors bars represent the standard deviation of the distribution. The average error from the 2000 estimates are plotted as the dotted lines above and below the best estimate. The solid line is the the value of $\mathcal{P}$ (i.e. input) at that frequency. The frequency error bars representing the width of the bin are omitted for clarity. The horizontal lines in the histogram plots represent the mean (solid) and input model (dotted). }
\label{fig:psd_hist_plot}
\end{figure}

In this work, we take the input power spectrum $\mathcal{P}$ to be a broken power-law of the form:
\begin{equation}
\mathcal{P}(f) = A_b \left(\frac{f}{f_b}\right)^{\alpha}
\end{equation}
where $\alpha=-1$ below some break frequency $f_b$, and $\alpha=-1.5$ above it, and $A_b$ is a normalization factor. We take $f_b=10^{-6}$ Hz. This break frequency is consistent with a black hole of $\sim5\times10^7 M_{\odot}$, which is typical of many Seyfert galaxies \citep{2006Natur.444..730M}. The lag in the simulated light curves is taken to be constant in phase (at 1 radians), so that the lag scales with $\tau(f)\propto f^{-1}$. For the gaps, we tried different patterns, as will be discussed. The most relevant given X-ray observations are those which have a period of $\sim 1.6$ hours, typical of low earth orbit observations. Throughout the following simulations, we study two cases: $A_b = 3\times10^6$ and $A_b = 8\times10^{3}$, representing high and low power cases respectively. The count rates for the two cases are 5 and 1 respectively. These are chosen as typical values for a bright variable, and relatively faint, less variable sources, corresponding to rms variability of $\sim$ 35 and 10 percent in each case. For each of these two cases, we run simulations with, and without gaps. The simulations without gaps are used for comparison and consistency check. All the simulated light curves are equivalent to an exposure of 200 ks. In simulations with gaps, we discuss both on-source and total exposures of 200 ks as detailed below.

\subsection{Power spectra}\label{sec:psd_sim}
 First, we discuss estimating the power spectrum, starting with a simple, high power, high signal to noise case without including gaps to use as a proof of concept. We simulated more than 2000 light curve realizations from case-1 PSD with 1 second sampling rate, we add Poisson noise then binned the light curves to 512 second bins. We used frequency bins that give in the case of even sampling at least 10 Fourier frequency points per bin. The same experiment is repeated for the low-rms case. The model PSDs and typical light curve realizations for these two cases are shown in Fig. \ref{fig:psds} and Fig. \ref{fig:lc}.

The result is summarized in Fig. \ref{fig:psd_hist_plot}, where we show the ensemble-averaged measured power at nine frequency bins, along with histogram distributions for two selected frequency bins. Each simulated light curve gives an estimate of the power spectrum at the nine frequencies and their uncertainties. The left panel of Fig. \ref{fig:psd_hist_plot} shows the means of these estimates (points). The errors on those points are taken to be the standard deviation of estimates around the mean. The average of the measured uncertainties are also plotted as the dotted lines around the best estimates. The right panel shows the distribution of the 2000 values for two selected frequency bins (1st and 5th bins).

There are several points to note from Fig. \ref{fig:psd_hist_plot}. The power spectrum is well-recovered both for the high and low rms cases. Even in noise-dominated bands in the low rms case ( $>2\times10^{-4}$ Hz, see Fig. \ref{fig:psds} ). The noise in the light curves is accounted for automatically ($\boldsymbol{n}$ and $\mathsf{C}_n$ in equations \ref{eqn:x_s_n} and \ref{eqn:def_Cx}), so that the measured PSD is the underlying, noise-less $\mathcal{P}$.The estimates are nearly Gaussian, particularly at intermediate frequencies. The distribution tails at the lowest frequencies are consistent with expectations from standard Fourier analysis for bins with a small number of averaged frequencies \citep[e.g.][]{1993MNRAS.261..612P}. The shape of the distribution depends essentially on the effective number of independent frequencies present in the light curve. The central limit theorem ensures that when a relatively large number is averaged, the distribution is Gaussian. If the number is small, an estimate is obtained, the errors may not be Gaussian, but the formalism presented here allow us to also estimate probability distribution of the lags using either direct evaluation of the likelihood function, or through Monte Carlo Markov Chains.

It is also clear that the method gives unbiased, consistent estimates of the power. The plot also shows that the uncertainty estimates (dotted envelopes), discussed in sec. \ref{sec:uncertainties} and taken here as the average of individual uncertainties, are very consistent with the standard deviation of an ensemble of estimates. In fact, for the cases of Fig. \ref{fig:psd_hist_plot} where no gaps are included, the frequencies are independent, and so the uncertainties taken directly from the Fisher matrix and those estimated by stepping through the parameter space are the same.

\begin{figure}
\centering
 \includegraphics[width=230pt,clip ]{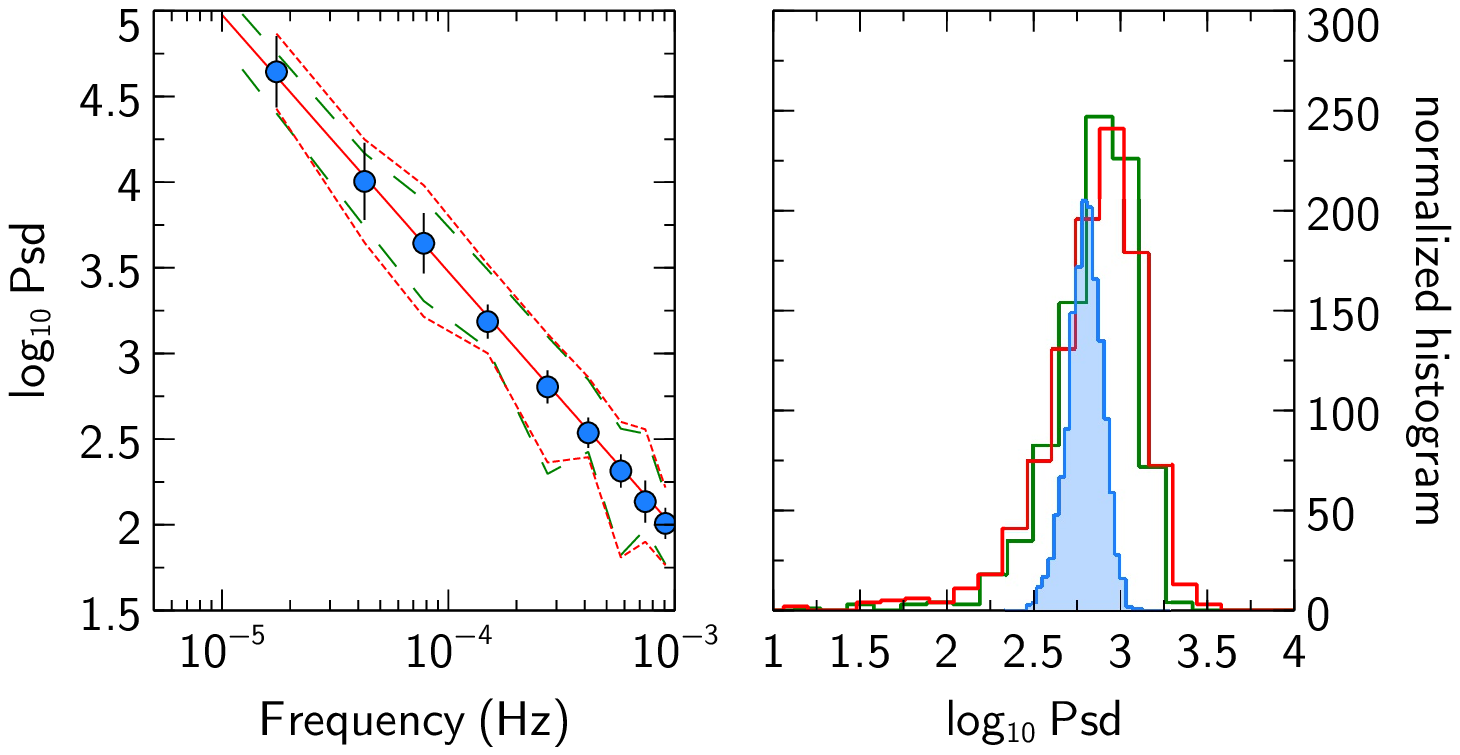}
 \includegraphics[width=230pt,clip ]{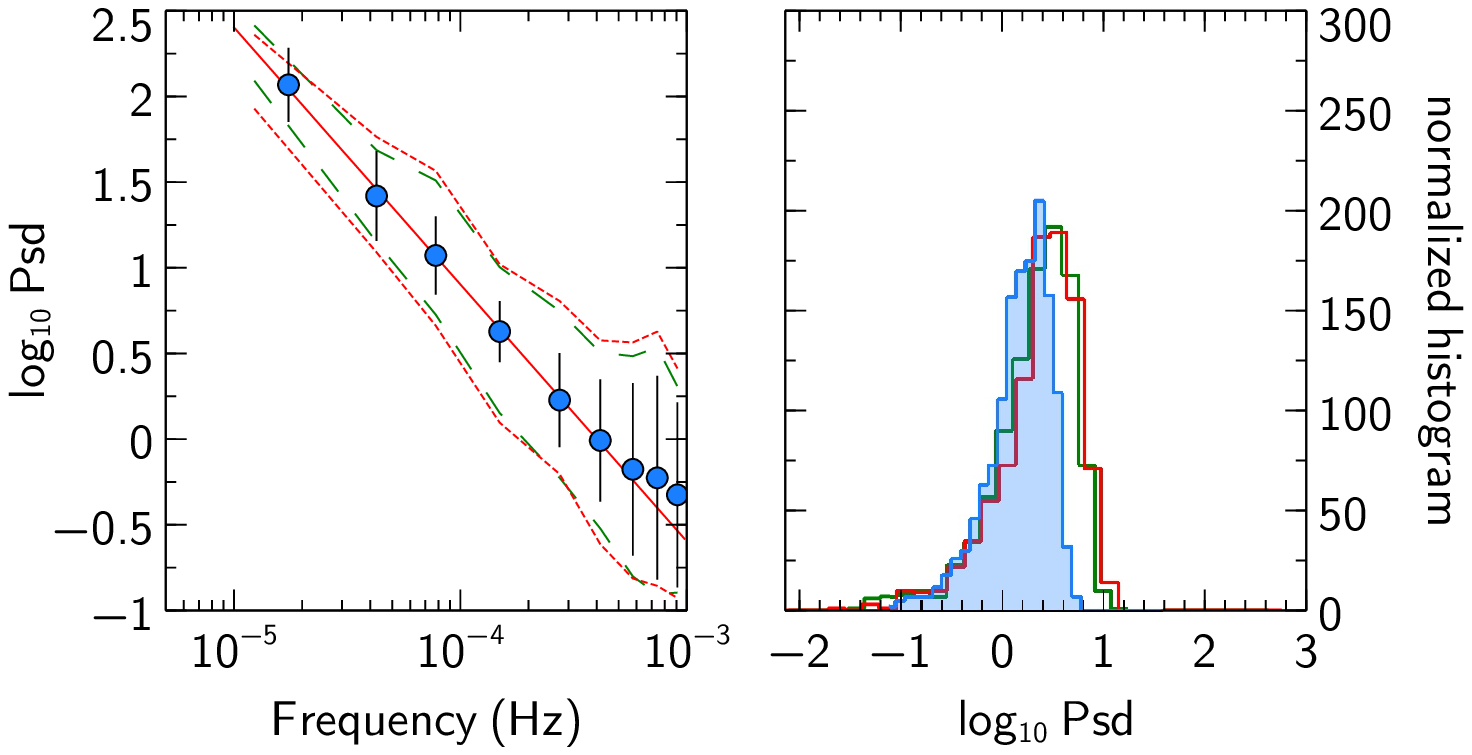}
\caption{{\it Left: } The average estimated power spectra (similar to Fig. \ref{fig:psd_hist_plot} ) including light curves with gaps for high (Top) and low (bottom) rms cases. The points represent the distribution means for the case of no gaps (similar to Fig. \ref{fig:psd_hist_plot}) and the error bars are the standard deviation around the mean. The red dotted envelope is the standard deviation of the estimates including gaps with the light curves of length 200 ks (data + gaps, G1). The green dashed envelope is the standard deviation of the estimates for light curves with gaps, but with {\it on-source} exposure of 200 ks (data only, G2). {\it Right} A histogram distribution of the estimates at the 5th frequency bin ($\sim 2\times10^{-4}$ Hz) which corresponds roughly to the periodicity of the gaps. The blue line is for light curves without gaps. The red line is for the case of total exposure of 200 ks, and the green line is for the case of on-source exposure of 200 ks.}
\label{fig:psd_gaps}
\end{figure} 

Similar simulations were performed for the low and high rms cases (as defined in section \ref{sec:sims} and Fig. \ref{fig:psds}) considering light curves with gaps. For comparison, we simulate light curves where the length of observation is 200 ks, and also light curves where the on-source exposure is 200 ks. The gaps are generated randomly assuming that both the length of data stream and gaps are Gaussian random variables with means 5700 and 4000 seconds respectively, and a standard deviation of 100 seconds. These gap patterns roughly resemble those usually encountered in \emph{Suzaku} observations and relevant to NuSTAR and AstroH. Other gap patterns have also been explored, and the conclusion are in general the same (with the obvious change of the frequencies affected.). The result is plotted in Fig. \ref{fig:psd_gaps}. 

High and low rms cases are plotted in the top and bottom panels respectively.  In each case, the left plot is similar to that in Fig. \ref{fig:psd_hist_plot}. The points and the errors bars are for the case with no gaps for comparison. The red-dotted and green-dashed lines are the envelope of the standard deviation of the PSD estimates for light curves with gaps and light curve lengths of 200 ks (hereafter case G1) and on-source exposure of 200 ks (hereafter case G2) respectively.

The gaps have several effects compared to the continuous case. The errors are in general larger because there is less data on the whole, except for the very lowest frequencies where the errors in G2 are smaller than the no-gaps case because the requirement of a on-source exposure of 200 ks means there is more low frequency data. Also, the errors for both G1 and G2 are larger for frequencies close to the gap periodicity. The reason is that information on those frequencies are missing because of the gaps. This is a general result that we found throughout the simulations, and it shows that the periodic gaps cause the uncertainties at the frequency corresponding to the gaps periodicity ($\sim 1\times10^{-4}$ Hz). G1 has about 60-70\% less exposure and its errors are slightly larger then G2 (the difference for a single frequency band is not huge, but all frequency bins are affected). The distribution histogram for the frequency bin closes to the gaps frequencies are also plotted in Fig. \ref{fig:psd_gaps}.

\subsection{Time lag}\label{sec:lag_sim}
\begin{figure}
\centering
 \includegraphics[width=230pt,clip ]{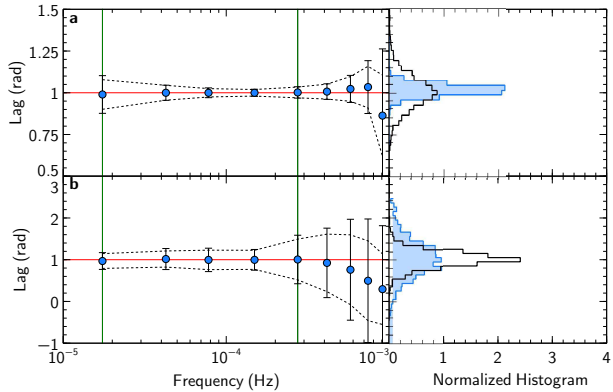}
\caption{Similar to Fig. \ref{fig:psd_hist_plot} but now showing lags instead of psd. High and low rms cases are shown in panels a (top) and b (bottom) respectively for light curves without gaps. The average estimated lag is shown as points. The standard deviation around the mean is shown in as error pars. The envelope dotted line shows the average estimated uncertainties. The right panels in each case show the (normalized) number of values histogram for the 1st (un-shaded) and 5th (shaded) frequency bins, marked with vertical lines in the left panels..}
\label{fig:lag_hist_plot}
\end{figure}

Analysis similar to that presented in section \ref{sec:psd_sim} was extended to include time lags. Fig. \ref{fig:lc} show typical light curves pairs for the high and low rms cases defined in section \ref{sec:sims}, where for each pair, the second light curve is shifted with a phase of 1 radian. The results are presented in Fig. \ref{fig:lag_hist_plot}. Although the presented simulations are for the case of constant phase lag at 1 radians, we tested for other forms (e.g. constant time lag, a time lag that has functional dependence on $f$ etc.), and the results are not different from those discussed here.

The figure, analogous to Fig. \ref{fig:psd_hist_plot}, shows the averaged lag calculated from an ensemble of 2000 light curve realization for the high and low rms cases without gaps. The lags are well-recovered for all frequencies for both cases. The distributions of the estimates (shown in Fig. \ref{fig:lag_hist_plot}) are almost perfect Gaussians. The plot also shows that the uncertainty estimates (dotted envelopes), discussed in sec. \ref{sec:uncertainties} and taken here as the ensemble average of individual uncertainties, are also consistent with the standard deviation of an ensemble of estimates. The slight difference at the noise-dominated frequencies (highest frequencies at panel b in Fig. \ref{fig:lag_hist_plot}) is an artifact of the simulation, where the noise-dominated parameters sometimes fail to converge, and it is therefore hard to obtain uncertainties and those are removed when estimating the average uncertainties. In practical data analysis, one would reduce the number of frequency bins to improve the signal to noise ratio.

\begin{figure}
\centering
 \includegraphics[width=230pt,clip ]{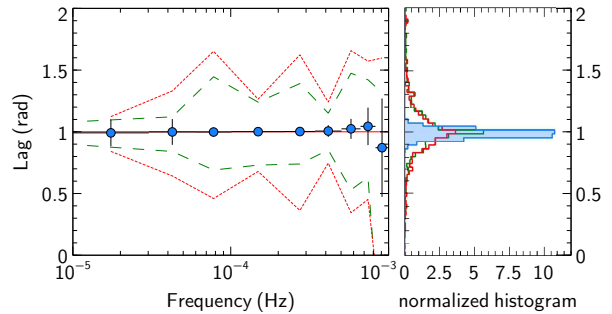}
\caption{Similar to Fig. \ref{fig:lag_hist_plot} but now including light curves with gaps for the high rms case. The average estimated lag for light curves without gaps is shown as points. The standard deviations around the mean are shown as error bars. The envelope dotted line (red) shows the standard deviation for light curves with gaps and length of 200 ks (G1). The envelope dashed line (green) is for the case of light curves with gaps and an on-source exposure of 200 ks (G2). The right panels shows the corresponding (normalized) number of values histogram for the 5th frequency bins, which corresponds roughly to the frequency of the periodic gaps.}
\label{fig:lag_hist_plot_gap3}
\end{figure}

Extending the analysis to light curves with gaps is again straight forward. Fig. \ref{fig:lag_hist_plot_gap3}. As in the case of power spectra, the errors are larger for light curves with gaps because less data is used. The lowest frequencies are not affected much because the long time-scale trends in the light curves are not affected if there are gaps on smaller time-scale. The periodic gaps have the effect of increasing the uncertainty of the measured lags at frequencies close to the gaps frequency and also its harmonics where information is missing. This, combined with the gap randomness (i.e. it is not a single frequency) and frequency binning produces the fluctuations seen in Fig.  \ref{fig:lag_hist_plot_gap3}. The results for the low rms case is very similar. The low signal to noise ratio however means the errors are larger, and sometimes simulations not constrained. Better estimates are obtained when using less frequency bins (i.e. improving the signal per bin), and in this case, the results are similar to those of the high rms case.

\begin{figure}
\centering
 \includegraphics[width=230pt,clip ]{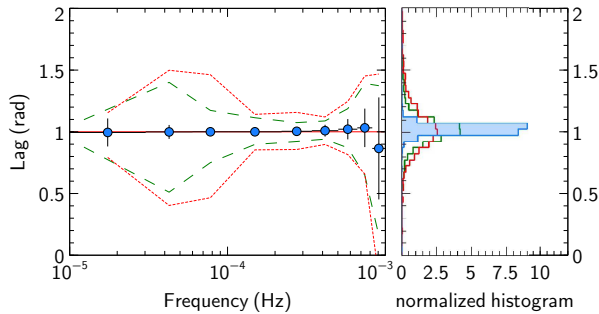}
\caption{Similar to Fig. \ref{fig:lag_hist_plot_gap3} but with a different gap pattern. The gaps now have a periodicity that corresponds to a frequency of $\sim 7\times10^{-5}$ Hz. The histograms are for the second frequency point.}
\label{fig:lag_hist_plot_gap}
\end{figure} 

This increased uncertainty at the gaps periodicity is further illustrated in Fig. \ref{fig:lag_hist_plot_gap}, which is similar to Fig. \ref{fig:lag_hist_plot_gap3} but for a different gap pattern. Here, the gaps have a periodicity corresponding to a frequency of $\sim 7\times10^{-5}$ Hz. Again, the effect of the gaps is that less information is available to the gap frequency, and therefore the uncertainty is larger. The distribution of the estimates is Gaussian or very close to Gaussian in most cases. The power of the likelihood method presented here is that, even in frequency bands where the effective number of independent frequencies is small, one can obtain a direct measure of the probability distribution whether by stepping through the likelihood function, or more efficiently by using MCMC.

\section{Applications}\label{sec:applications}
In this section, we discuss the application of the above method to calculate time lags in \emph{Suzaku} observations of two sources: NGC 4151 and MCG-5-23-16. Time delays have been seen in this two objects using XMM-Newton and standard lag calculation methods \citep{2012MNRAS.422..129Z,2013ApJ...767..121Z}. The observations used in the following discussion are summarized in table 1.

\begin{table}
\begin{tabular}{llll}
\hline
Object  &  Obs. ID & Exposure (ks) & Date \\
\hline
NGC 4151 & 701034010 & 125 & 08-2008 \\
		 & 906006010 & 60  & 04-2011 \\
		 & 906006020 & 60  & 04-2001 \\
MCG-5-23-16 & 700002010 & 95 & 12-2005\\ \hline
\end{tabular}
\caption{Observations of NGC 4151 and MCG-5-23-16 from the \emph{Suzaku} archive used in this work}
\end{table}

Data were retrieved from the archives and reduced using \textsc{heasoft 6.13} and the latest calibration files (\textsc{caldb} version 20130305). Cleaned events files for all XIS detectors operational during the observations were produced following the \emph{Suzaku} user guide. Then \textsc{xselect} was used to extract source and background light curves with time bins of 512 seconds. The source region in each case was circular with radius of 3.5 arcmin, and background regions were selected from source-free regions on the CCD. In order to study time lags, we extracted light curves in eight energy bins between $2-10$ keV in steps of 1 keV. Background light curves were than scaled to match the area of the source region before subtracting them from the source light curves. The XIS0 and XIS3 counts were combined to produce a total front-illuminated light curve, while the XIS1 gives a back-illuminated light curve. The two light curves can then be fitted simultaneously using the formalism discussed in section \ref{sec:likelihood}.

\begin{figure*}
\centering
\begin{tabular}{cc}
 \includegraphics[width=210pt,clip ]{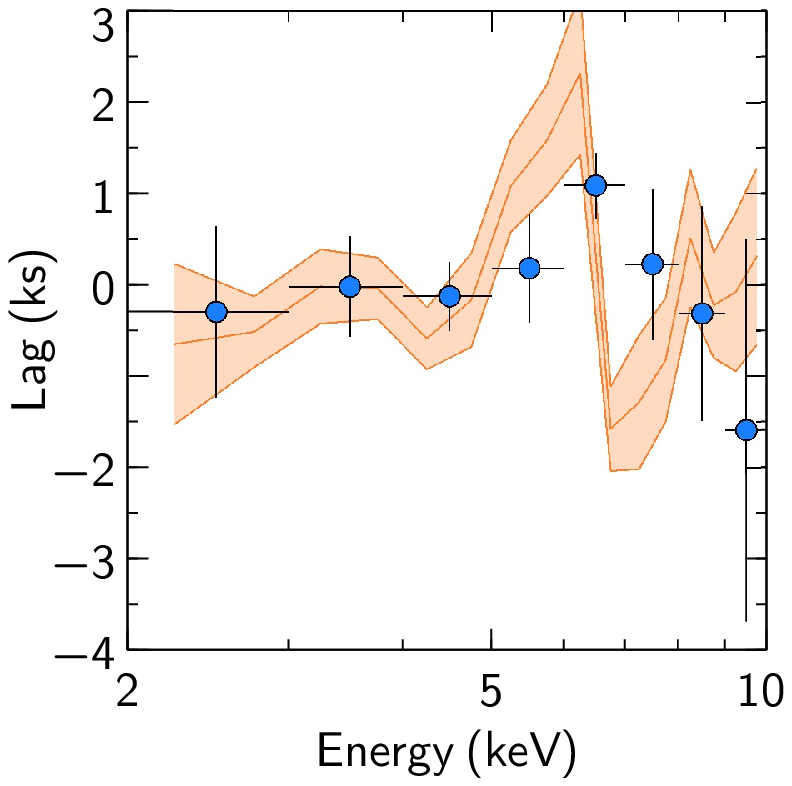}&
 \includegraphics[width=210pt,clip ]{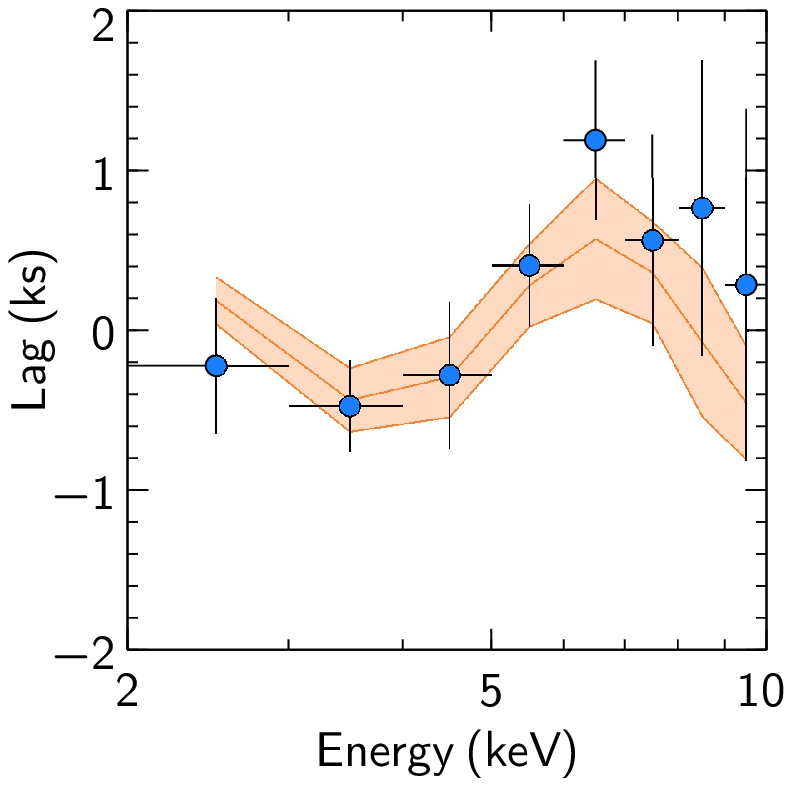}
 \end{tabular}
\caption{Lag-energy plot for NGC 4151 (\emph{left}) and MCG-5-23-16 (\emph{right}). The points represent the estimated lag at frequencies $3\times10^{-5}-3\times10^{-4}$ Hz (NGC 4151) and $10^{-5}-3\times10^{-4}$ Hz (MCG-5-23-16) between the marked energy band and the the whole 2-10 keV band taken as a reference (after removing the band of interest to keep the noise uncorrelated. See \citealt{2012MNRAS.422..129Z} for details on lag-energy plots). The shaded plots are the lag-energy plots from the XMM-Newton data \citep{2012MNRAS.422..129Z,2013ApJ...767..121Z} }
\label{fig:data}
\end{figure*} 

Fig. \ref{fig:data} shows the lag-energy plots for NGC 4151 (left) and MCG-5-23-16 (right), along with plots from previously published XMM-Newton data \citep{2012MNRAS.422..129Z,2013ApJ...767..121Z}. Each point in the plots is a result of maximizing the likelihood function for the power spectra and phase lags between individual light curves and the total $2-10$ keV light curves, excluding the current energies \citep[see][ for details on lag-energy plots]{2012MNRAS.422..129Z}. The plotted frequencies are $3\times10^{-5}-3\times10^{-4}$ Hz and $10^{-5}-3\times10^{-4}$ Hz respectively. Although the uncertainties at the highest energies are relatively large, it is clear that there is a structure at $6-7$ keV consistent that seen in the XMM-Newton data. The match between the \emph{Suzaku} and XMM-Newton plots in the case of MCG-5-23-16 (Fig. \ref{fig:data}-right) is remarkable. For the case of NGC 4151, although the shapes are statistically consistent, the lag-energy shape in this source is known to be flux- and frequency-dependent \citep{2012MNRAS.422..129Z}. The length and quality of the \emph{Suzaku} observations do not allow direct comparison at the same exact frequencies, but the fact that there is a peak at $\sim 6$ keV adds further evidence that the iron line is responsible for these lags. A further test is achieved by adding artificial gaps to the XMM-Newton and calculating lags. This however reduces the amount of available data and smears the signals out.

\begin{figure}
\centering
\begin{tabular}{cc}
 \includegraphics[width=180pt,clip ]{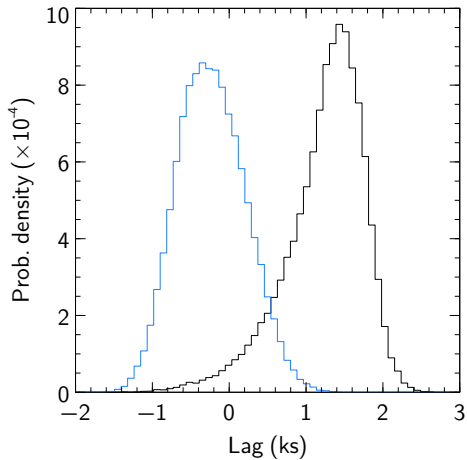}&
 \end{tabular}
\caption{Probability densities for the MCG-5-23-16 two lag points at $1-2$ and $6-7$ keV shown in Fig. \ref{fig:data} estimated using Monte Carlo Markov Chain.}
\label{fig:mcg_prob}
\end{figure} 

The likelihood method allows us to obtain full probability densities for lag estimates and hence quantify directly the significance of any lag detection. For example, Fig. \ref{fig:mcg_prob} shows the probability density of the estimated lag values at $1-2$ and $6-7$ keV for the case of MCG-5-23-16, plotted in the right panel of Fig. \ref{fig:data}. After the best estimates are obtained by maximizing the likelihood function, the multi-dimensional parameter space is mapped out using MCMC. The chains were generated with an affine-invariant ensemble sampler \citep{STMAZ.05709093,2013PASP..125..306F}, and the result is a probability density for each of the estimated values.

\section{Summary}
We presented and discussed a method to calculate frequency-dependent power spectra and time lags for unevenly-sampled data. The method, first introduced by \cite{2010MNRAS.403..196M}, relies on likelihood maximization, and gives the most likely power and lag estimates given the data. We tested the method using Monte Carlo simulations, and showed that the main effect of periodic gaps, typical of low-earth orbit X-ray observations, is to give unconstrained estimates at the frequency corresponding to the gap periodicity, while information at other frequencies is recovered. We applied the method to \emph{Suzaku} archival observations of NGC 4151 and MCG-5-23-16, and showed that their lag-energy spectra are consistent with those observed using XMM-Newton, giving further support to their interpretations as being due to relativistic reverberation close to the black holes.

\bibliographystyle{astron}
\bibliography{bibliography}

\end{document}